\UseRawInputEncoding
\documentclass[10pt, a4paper]{article}
\pdfoutput=1
\usepackage[final]{lrec-coling2024} 
\usepackage{multirow}
\usepackage{graphicx}
\graphicspath{{picture/}}
\usepackage[normalem]{ulem}
\useunder{\uline}{\ul}{}
\usepackage{enumitem}
\PassOptionsToPackage{table,xcdraw}{xcolor}
\usepackage{bm}
 \usepackage{verbatim} 
 \usepackage{amssymb}

\title{\vspace*{.5\baselineskip} \textbf{Persona-centric Metamorphic Relation  guided Robustness Evaluation for Multi-turn Dialogue Modelling}}

\name{Yanbing Chen\textsuperscript{1}, Lin Li\textsuperscript{1}, Xiaohui Tao\textsuperscript{2} , Dong Zhou\textsuperscript{3}} 

\address{\textsuperscript{1}Wuhan University of Technology,\\ \textsuperscript{2}University of Southern Queensland, \\\textsuperscript{3}Guangdong University of Foreign Studies \\
         chenyb@whut.edu.cn, cathylilin@whut.edu.cn, 	Xiaohui.Tao@unisq.edu.au,\\dongzhou@gdufs.edu.cn\\
         }

\abstract{
Recently there has been significant progress in the field of dialogue system thanks to the introduction of training paradigms such as fine-tune and prompt learning. Persona can function as the prior knowledge for maintaining the personality consistency of dialogue systems, which makes it perform well on accuracy. Nonetheless, the conventional reference-based evaluation method falls short in capturing the genuine text comprehension prowess of the model, significantly relying on the quality of data annotation. In contrast, the application of metamorphic testing offers a more profound insight into the model's distinct capabilities without necessitating supplementary annotation labels. This approach furnishes a more comprehensive portrayal of the model's intricacies and exposes intricacies concealed within reference-based validation techniques. Consequently, we introduce a persona-centric metamorphic relation construction for metamorphic testing, aimed at evaluating both the persona consistency and robustness of personalized dialogue models. For that reason, this work evaluates several widely used training paradigms including learning from scratch, pretrain + fine-tune and prompt learning in personalized dialogue retrieval to know if they are more robust or if they have the same flaws as their predecessor. Under three kinds of designed metamorphic relations with consistent outputs, our experimental results reveal that prompt learning shows stronger robustness compared to training from scratch and fine-tune. Although tested retrieval models gain competitively high retrieval accuracy according to the traditional reference-based validation, they are still fragile and demonstrate various unexpected behaviors, thus there is still room for future improvement in personalized dialogue retrieval.
 \\ \newline  }

\begin{document}

\maketitleabstract

\section{Introduction}
The fine-tune and prompt learning paradigms \cite{prompt_survey} have made remarkable progress in tackling challenges related to dialogue systems, including dialogue retrieval \cite{TransferTransfo,Poly_encoders}, generation \cite{25,26}, and summarization \cite{27,28}. 
Within the dialogue system, retrieval-based response selection plays a pivotal role. It not only synergizes effectively with other modules, such as the generative module, but also exhibits impressive standalone performance.  
Among these advancements, the integration of prior knowledge, such as persona \cite{persona-chat_dataset,5,DIM,29,45} and background information \cite{FIRE,30,44,46}, into the dialogue retrieval system stands out as a significant breakthrough. 
This incorporation enhances retrieval accuracy and introduces a personalized dialogue retrieval \cite{3,BERT-CRA,COLING}. In this task, the system selects an appropriate response from a set of candidates based on the context and the speaker's persona. In the realm of multi-turn dialogue models, maintaining personality consistency is often a challenge due to the diverse nature of training data comprising different speakers and the absence of explicit long-term memory for each speaker. However, by incorporating detailed portrait information of the speaker's persona, the dialogue retrieval system effectively ensures consistency in personality, leading to substantial improvements in accuracy and overall system performance.

Although significant advances have been made in multi-turn dialogue retrieval tasks by incorporating persona and proposing various optimization algorithms to enhance dialogue models \cite{BERT-CRA,COLING,31}, the validation methods for these models have not received adequate attention. Currently, most evaluation of dialogue models still relies on reference-based methods, where the model's outputs are compared with ground truth samples to assess consistency. 
However, this widely adopted approach has practical limitations. {\bf{Firstly, its dependence on extensively annotated labels poses substantial demands.}} For generative dialogues, human evaluations of generated content entail individuals possessing both knowledge and discernment. In cases of retrieval dialogues, meticulously curated datasets are requisite. These demands translate to labor-intensive efforts, with limited assessments potentially failing to unveil real-world application issues effectively. {\bf{Secondly, the consistency achieved through reference-based evaluation may not result from a genuine understanding of the underlying text and logical reasoning}}, but rather from the model memorizing specific expression patterns or keywords \cite{Stress_Test,Beyond_Accuracy}. Moreover, this approach provides only an overall performance assessment of the model, without revealing specific strengths and weaknesses\cite{wuda_MT}. {\bf{Lastly, in the case of personalized dialogue systems, the failure to address the system's shortcomings in maintaining a consistent personality can lead to user perception of the system as untrustworthy}}, eroding confidence in its real-world capabilities. 

To tackle this pervasive challenge in the field of natural language processing, researchers have recently turned to metamorphic testing (MT), a technique commonly employed in software engineering. Metamorphic testing breaks the reliance on annotation labels and shifts focus to comparing the relationships between multiple inputs and outputs using a predefined set of attributes known as metamorphic relations (MR). This approach has been successfully applied to tasks such as sentiment analysis, machine translation, natural language inference, and question answering \cite{Systematicity_Perspective,Beyond_Accuracy,20,21}. 
However, it is worth noting that recent models designed for personalized dialogue retrieval have not undergone systematic testing to demonstrate their robustness under perturbations from persona-based metamorphic relations, where consistent outputs (Details in Sec.~\ref{meta}). Therefore, a validation method is needed to assess the persona-centric language comprehension of these personalized dialogue models and uncover their specific strengths and weaknesses. 

In this study, we assess the persona-based robustness of various personalized dialogue retrieval models. Our objective is to examine the capacity of dialog systems, employing various architectures, to maintain consistent personality in responses while being exposed to different persona perturbations. 
Specifically, we categorize personalized dialogue retrieval models into three training paradigms: non-pretraining techniques like Bi-LSTM \cite{LSTM}, fine-tune approaches based on BERT \cite{BERT}, and the increasingly successful paradigm of prompt learning \cite{prompt_survey}. Additionally, we define the perturbation-sensitive property of response to persona in personalized dialogues as a metamorphic relation.
Afterward, we conducted metamorphic testings to complement the reference-based validation approach by more specifically assessing the robustness of each model against persona-centric metamorphic relations, akin to those investigated in prior studies that employed equivalence relations \cite{Systematicity_Perspective}. 
\begin{figure}[ht]
\begin{center}
\setlength{\abovecaptionskip}{0cm}
  \centering
  \includegraphics[width=0.9\linewidth,page={1}]{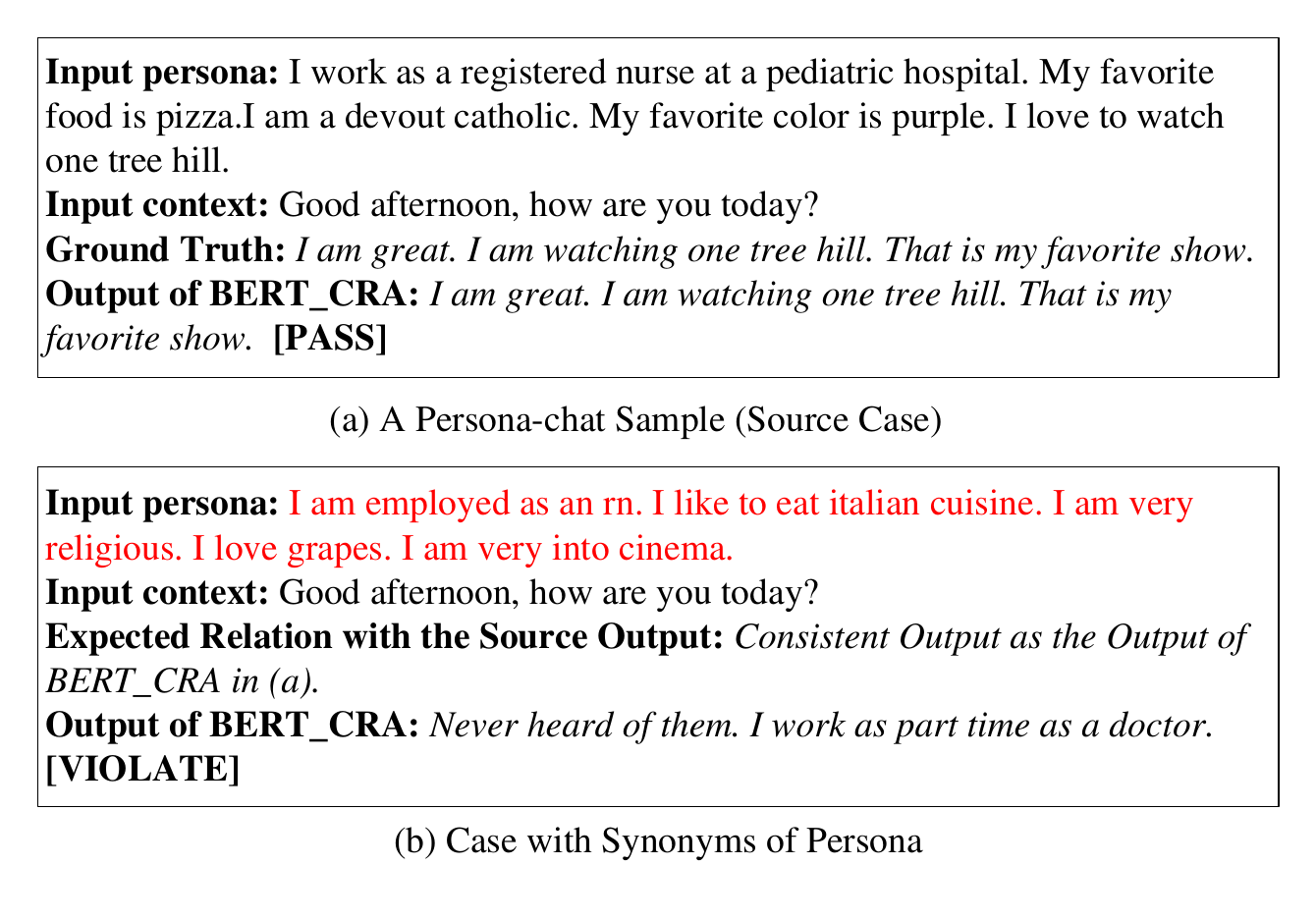}
  \caption{References-based Validation}
  \label{violation text}
  \vspace{-0.5cm}   
\end{center}
\end{figure}
These tests involved introducing noise and replacing synonyms \cite{17} to evaluate the models' performance. We examined various scenarios, including noise perturbations of character-level exchanging, sentence transformations using synonymous phrases, and variations in partners' personas. For instance, we perturbed each word of the persona by randomly swapping two characters, resulting in variations such as [\emph{love the music} → \emph{lvoe teh muisc}], and so forth. 

Our main contributions are summarized as follows: 1) three metamorphic relations based on the persona are designed, aiming to ensure personality consistency and highlight the conversational model's robustness; 2) to support our claims, a comprehensive quantitative analysis shows that prompt learning is stronger robustness compared to training from scratch and fine-tune. Although tested retrieval models gain competitively high retrieval accuracy according to the traditional reference-based validation, our persona-centric metamorphic testings have an impact on all retrieval models, albeit with varying degrees.

The dataset, replication package, and detailed results for this paper will be made available online.

\section{Limitation on Reference-Based Validation} \label{limitation}
In the personalized dialogue retrieval task, the effectiveness of popular methods can be easily assessed by comparing the model's output with the true labels of the samples. High accuracy indicates excellent task performance for these models. The reference-based method used to evaluate personalized dialogue systems has practical limitations, despite its widespread adoption.

\begin{table*}[htb]
\setlength{\abovecaptionskip}{0.2cm}   
\centering 
\tiny
    \resizebox{9.5cm}{!}{
            \begin{tabular}{@{}ll@{}}
            \hline
            \multicolumn{1}{c|} {Self-persona} & 
            \multicolumn{1}{c}{Partner-persona} \\ 
            \hline
            \multicolumn{1}{l|}{
            \begin{tabular}[c]{@{}l@{}}
                \color[HTML]{009901}I am afraid of heights.\\ 
                I love animals and have two dogs.\\ 
                I am Native American and live in Oklahoma.\\ 
                I work as an attorney.\\ 
                I am not religious.
            \end{tabular}}                                                
            & \begin{tabular}[c]{@{}l@{}}
                I love watching movies and TV.\\ 
                I have a husband who I love very much.\\ 
                I do not like exercise or physical activity.\\ 
                My favorite food is ice cream.\\ 
                \color[HTML]{3531FF}I am a homebody.
            \end{tabular}                                              \\ 
            \hline
            \multicolumn{2}{l}{
            \begin{tabular}[c]{@{}l@{}}
            Partner\ \,: Hey there. How are you?\\ 
            Self\qquad: Good, do you like animals?\\ 
            Partner\ \,: Yeah, I like cats. I have one.\\ 
            Self\qquad: I have 2 dogs, they are great, where do you work?\\ 
            Partner\ \,: \color[HTML]{3531FF}I stay at home with the kids.\\ 
            Self\qquad: \color[HTML]{009901}Are you afraid of heights? I certainly am.\\ 
            Partner\ \,: No. Do you like TV?\\ 
            Self\qquad: Sure, I like TV, what do you watch?\\ 
            Partner\ \,: Really anything, what about you?\\ 
            Self\qquad: I do not have time to watch TV, I am an attorney, so I work a lot.\\ 
            Partner\ \,: I am not a very active person.\\ 
            Self\qquad: I certainly am, I am part Native American, I live here in Oklahoma.\\ 
            Partner\ \,: Oh. Tell me something about yourself.\\ 
            Self\qquad: Well, I do not like heights very much and I love animals. What about you?\\ 
            Partner\ \,: \color[HTML]{FE0000}I am a boring person.\\ 
            Self\qquad: \color[HTML]{FE0000}I am not much fun either. So what else is new with you?   
            \end{tabular}}\\ 
            \hline
        \end{tabular}
        }
      \caption{Persona-chat dataset example. The green denotes sentences that demonstrate self-persona, the blue represents sentences that exemplify partner-persona, and the red represents the context and response of the last turn of the complete conversation.}
    \label{tab:persona_chat_dataset}
    \vspace{-0.3cm}   
\end{table*}

\subsection{Compulsory Labels}
In the case of retrieval dialogues, the requirement for meticulously curated datasets is crucial. These datasets serve as references for evaluating the system's ability to retrieve relevant responses. The process of creating such a dataset necessitates extensive human effort and the application of relevant knowledge to meticulously choose and annotate suitable responses that align with a given context or query. It involves a significant investment of human labor and requires individuals with expertise in the field to ensure the careful curation of the dataset. The meticulous curation process further adds to the labor-intensive nature of evaluating personalized retrieval-based dialogue systems. Additionally, the constraints of human resources and time can lead to limited assessments, which might not fully capture the complexities and challenges that dialogue systems may encounter in applications.
\subsection{Limitation on Capability with Reference-Based Validation}
The examples as illustrated in Figure \ref{violation text}, reveal an important limitation. When we modify the persona to a synonymous sentence with a similar meaning and input it into the model, it becomes confused, resulting in an output that deviates from the expected relationship with the original source output. This indicates a lack of robustness in the model's performance. Moreover, some researches find that certain models may achieve correct answers through the memorization of specific patterns or keywords rather than a genuine understanding and inference of the answers \cite{40}. 

Traditional reference-based validation methods solely focus on reporting the "consistency" of actual outputs with the true labels, and unfortunately, they do not effectively reveal the true language comprehension capabilities of the models \cite{Beyond_Accuracy}. Additionally, these methods fail to capture the associated shortcomings of model robustness. While traditional reference-based metrics provide a basic understanding of model performance, there is a need for more comprehensive and rigorous evaluation methods, such as metamorphic testing, to assess the consistency and robustness of personalized dialogue retrieval systems under challenging conditions. This can help identify potential weaknesses and limitations of existing methods and guide the development of more reliable and robust personalized dialogue systems.

\section{TASK Description}

\subsection{Task}
In the domain of personalized multi-turn dialogue systems, various types of datasets exist, such as personalized datasets \cite{48}, personalized empathy datasets \cite{PEC}, and datasets incorporating personalized information with background knowledge \cite{49}. For this paper, our primary focus is on the Persona-Chat dataset \cite{persona-chat_dataset}, while acknowledging the presence of other relevant datasets that will be explored and discussed in future studies.

The Persona-Chat dataset, represented as $D$, consisting of n conversation tuples in the format of $ (c,p,r,y) $. Specifically, $ c=\left \{ u_{1}, u_{2},..., u_{n_{c}} \right \} $  represents the $n_{c}$ context utterances, $ p=\left \{ p_{1}, p_{2},..., p_{n_{p}} \right \}$  is the $n_{p}$ personas of the speaker, and $r$ is the response candidate for $c$. The label $ y \in \left \{ 0, 1\right \} $ indicates whether $r$ is the appropriate response for $ (c, p) $, where $ y = 1 $ means it is appropriate, while $ y = 0 $ means it is not. Our objective is to learn a matching function $g$ from $ D(c,p,r) $ such that, given any tuple $(c,p,r)$, $g(c,p,r)$ calculates the degree of matching between $(c,p)$ and $r$.

The Persona-Chat dataset \cite{persona-chat_dataset} is currently the most extensive publicly available dataset of multi-turn dialogues conditioned on persona. It consists of 65,719 context-response pairs for the training set, 7,801 for the validation set, and 7,512 for the test set. The dataset includes correct responses from real humans, while incorrect responses are randomly sampled. To increase the challenge of the task, measures are taken to ensure that there is no overlap of contexts and roles between the training, validation, and test sets. This approach guarantees a robust evaluation of models' generalization capabilities across diverse scenarios. Table \ref{tab:persona_chat_dataset} presents a selection of illustrative examples from this diverse dataset. Several studies \cite{DIM,FIRE,BERT-CRA,COLING,TransferTransfo} have explored both non-pre-trained and pre-trained approaches on this dataset to enhance personalized dialogue retrieval and maintain consistent personality in the responses.

\subsection{Tested Models}
\begin{figure}[ht]
  \centering
  \setlength{\abovecaptionskip}{0cm}
  \includegraphics[width=1\linewidth,page={1}]{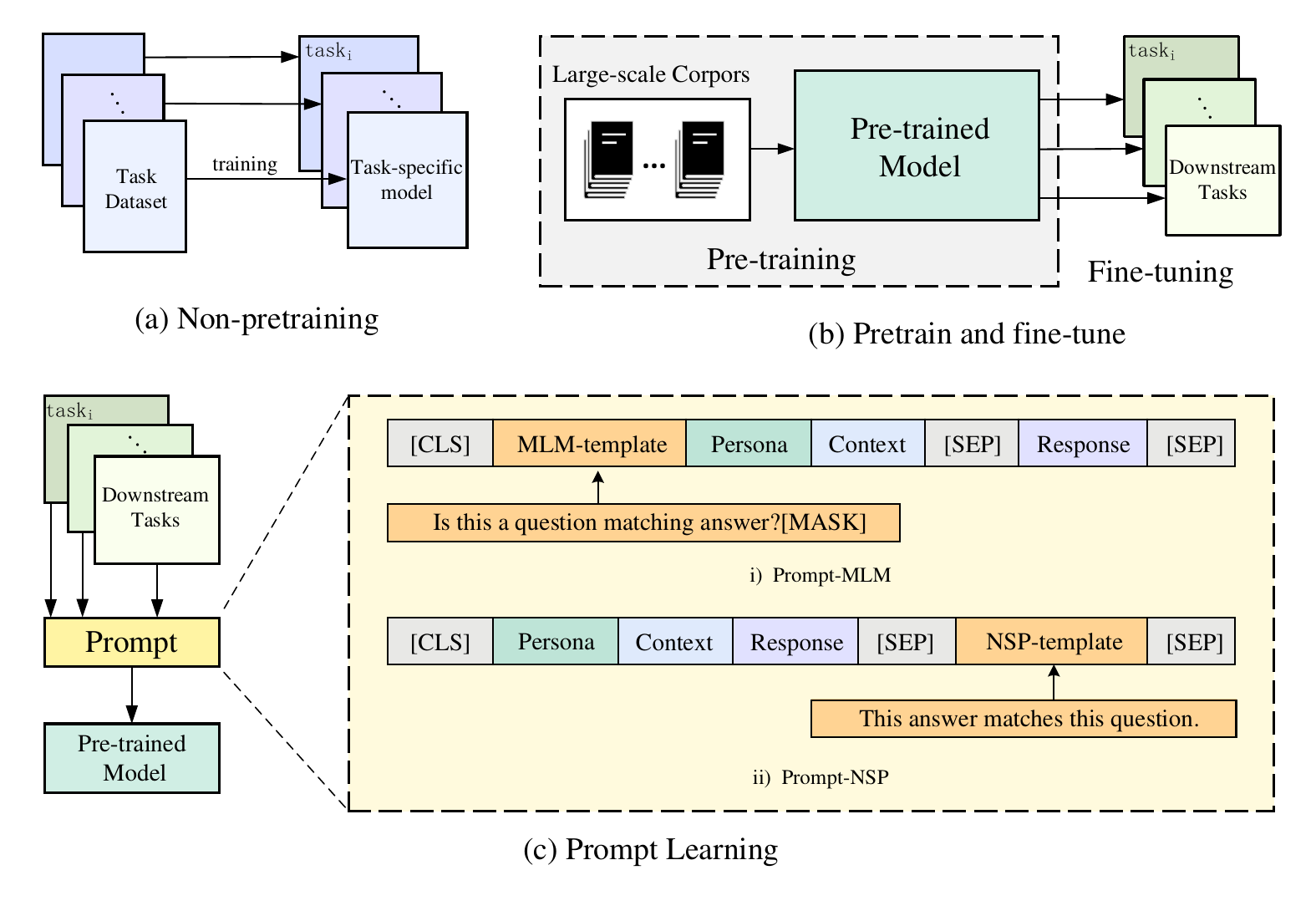}
  \caption{Three Training Paradigms}
  \label{Three Paradigm}
  \vspace{-0.2cm}
\end{figure}

In our study of multi-turn personalized dialogue retrieval, we investigate three paradigms and conduct evaluations of the task's robustness performance.
As shown in Figure~\ref{Three Paradigm}(a), the non-pretraining approach involves training the personalized dialogue retrieval model using dataset $D(c,p,r,y)$, such as \textbf{DIM}~\cite{DIM} and \textbf{FIRE}~\cite{FIRE}.

Pretraining methods utilize a pretrained model trained on a large-scale corpus. Fine-tuning initializes the personalized dialogue retrieval model with pretrained model parameters and updates its weights using dataset $D$ and matching function $g$ to adapt it to the task, as shown in Figure~\ref{Three Paradigm}(b). We will test \textbf{CoBERT}~\cite{PEC} and \textbf{BERT-CRA}~\cite{BERT-CRA}.

Illustrated in Figure~\ref{Three Paradigm}(c), the prompt learning includes an appropriate prompt $p$ within dataset $D$, following BERT's Masked Language Modeling (MLM) and Next Sentence Prediction (NSP) format. 
These prompts, such as \emph{cloze prompts} \cite{prompt_MLM} (\textbf{\romannumeral1)}) and prompts introduced by \citet{NSP}(\textbf{\romannumeral2)}), aid the pre-trained model in assimilating relevant information, improving personalized dialogue retrieval performance by leveraging acquired knowledge. 
Therefore,we predominantly utilize three models:\textbf{prompt-MLM(BERT)}, which is based on BERT-CRA converted to MLM format, \textbf{prompt-MLM(DialogLM)},which is based on the post-train dialogue model DialogLM\cite{DialogLM} and converted to MLM format, and \textbf{prompt-NSP(BERT)}, which is converted to NSP format. 

\section{Metamorphic Relations with Consistent Outputs} \label{meta}
The Persona-Chat(PC) dataset encompasses diverse persona sources, including \emph{self-persona}, \emph{partner-persona}, and scenarios involving both personas. It also considers cases such as \emph{no persona}, \emph{both personas}, or one of the two. Furthermore, the dataset incorporates a \emph{revised} persona description, which introduces complexity and tests the model's ability to generate accurate and contextually appropriate responses by reintroducing, promoting, or specializing original operations. To comprehensively investigate the robustness of persona-centric multi-turn dialogue models, we define the following metamorphic relations (MRs) to explore consistent outputs from three distinct viewpoints. 

    \textbf{MR1: Consistency with synonyms of persona} \label{MR1}
    Previous research \cite{Beyond_Accuracy,wuda_MT,Assessing} has employed the technique of replacing keywords or adjectives in a sentence with their respective synonyms to assess the model's comprehension of crucial aspects within the sentence. This approach aims to evaluate the model's understanding rather than solely focusing on improving accuracy by selecting words that are strictly similar in literal terms.
    Considering the impact of persona sentences, it becomes crucial to capture overall meaning rather than individual words. To address this, we propose substituting synonymous sentences for the input persona, expecting that this modification will not affect the prediction results of the original model.
    
\textbf{MT1: Synonymous Sentences Test}
In this test, we used MR1, where synonymous sentences replaced the \emph{original} persona in the PC dataset. The aim was to assess the model's ability to recognize synonymous sentences and adhere to persona-centric metamorphic relations, specifically invariance. We found that the \emph{revised} version of the PC dataset is well-suited for our testing purposes. Therefore, we directly used this test set to evaluate perturbations on the \emph{original self-persona} model.

Example 1 as an illustration of this metamorphic relation (MR). Since the two personas have identical meanings, any deviation in the model's output for these inputs would be considered an error.

\begin{figure}[ht]
\vspace{-0.6cm}   
\setlength{\abovecaptionskip}{0cm}
  \centering
  \includegraphics[width=0.85\linewidth,page={2}]{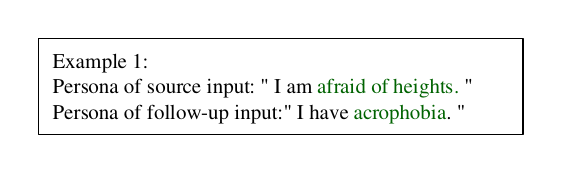}
  \label{MT1}
  \vspace{-0.6cm}   
\end{figure}
    
   \textbf{MR2: Consistency with persona of the partner interference}
    In a dialogue scenario, the model receives persona information from both participants simultaneously, including \emph{self} and \emph{partner}. As the conversation progresses in a multi-turn dialogue, the topics discussed may be relevant to both participants \cite{BERT-CRA}. Hence, it is crucial to consider the influence of the partner's persona on the subsequent conversation topics. Inferring the personas of both parties based on the history of multi-turn dialogues, in order to generate an appropriate response unaffected by external influences, is a critical aspect in assessing the model's robustness. To address this, we propose substituting the input persona with an unrelated \emph{partner-persona}. The expected outcome following this interference should align with the model's prediction prior to the intervention. Through such evaluations, we gain valuable insights into the model's ability to handle diverse persona inputs and assess its resilience in generating contextually appropriate responses.

\textbf{MT2: Persona Sensitive Test}
In the persona sensitive test, we replaced the original persona (e.g.,\emph{self-persona}) with the conversation partner's persona using MR2. Similar to the synonymous sentences test, we evaluated the \emph{self-persona} model using the \emph{partner-persona} version of the \emph{original} Persona-Chat dataset as our test set. Example 2 illustrates this transformation.

\begin{figure}[ht]
\vspace{-0.7cm}   
\setlength{\abovecaptionskip}{0cm}
  \centering
  \includegraphics[width=0.85\linewidth,page={3}]{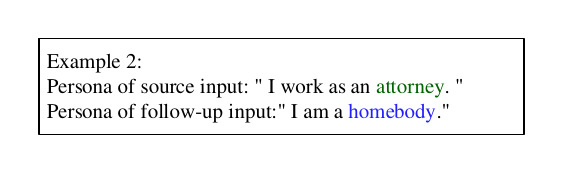}
  \label{MT2}
  \vspace{-0.7cm}   
\end{figure}

    \textbf{MR3: Consistent with character-level noise}
    Similar to previous works \cite{Beyond_Accuracy,Stress_Test}, our objective is to assess the models' robustness in realistic scenarios by introducing data perturbations. We inject noise, such as spelling errors, into the input data (e.g., persona) to uphold response consistency. To examine the model's performance under challenging conditions, we introduce character-level noise, simulating extreme cases. A robust model should maintain consistent predictions even with added noise. This evaluation provides valuable insights into the model's ability to handle variations and disturbances in input data.
    
    \textbf{MT3: Noise Test}
In contrast to the previous two MTs, MT3 assesses the \emph{original} versions of both the \emph{self-persona} model and the \emph{partner-persona} model, breaking away from relying on the transformed version provided in the PC dataset. To align with MR3, we redesigned the test set and implemented Swap Noise settings as described in \cite{Stress_Test}, which involves swapping one randomly selected pair of consecutive characters for each word in the text (e.g.,\emph{change} → \emph{chnage}). These noise experiments focus on the most recent-turn context of a multi-turn dialogue to gain a deeper understanding of the importance of persona and context in personalized dialogue retrieval. Examples 3-1, 3-2, and 3-3 illustrate the transformations for this metamorphic testing.

    \textbf{MT3-1: Persona-Only Test.}
    We limit the character exchange to randomly selected consecutive characters in each persona word.
    
        \begin{figure}[ht]
        \vspace{-0.7cm}   
        \setlength{\abovecaptionskip}{0cm}
          \centering
          \includegraphics[width=0.85\linewidth,page={4}]{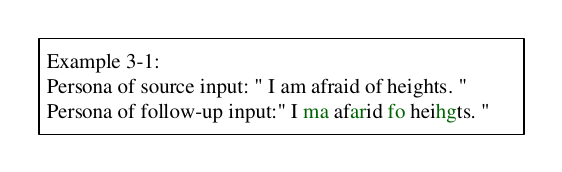}
          \label{MT3-1}
          \vspace{-0.7cm}   
        \end{figure}
    \textbf{MT3-2: Context-Only Test.}
    We only substitute randomly chosen consecutive characters in each context word.
    
        \begin{figure}[ht]
        \vspace{-0.7cm}   
        \setlength{\abovecaptionskip}{0cm}
          \centering
          \includegraphics[width=0.85\linewidth,page={5}]{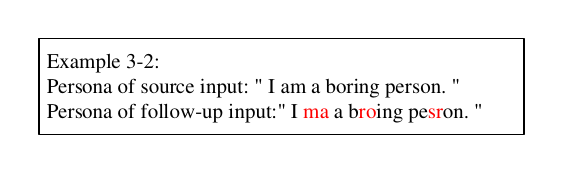}
          \label{MT3-2}
          \vspace{-0.7cm}   
        \end{figure}
    \textbf{MT3-3: Persona-and-Context Test.}
    We perform a simultaneous substitution of random consecutive characters for each word of both the persona and context.
        \begin{figure}[ht]
        \vspace{-0.7cm}   
        \setlength{\abovecaptionskip}{0cm}
          \centering
          \includegraphics[width=0.85\linewidth,page={6}]{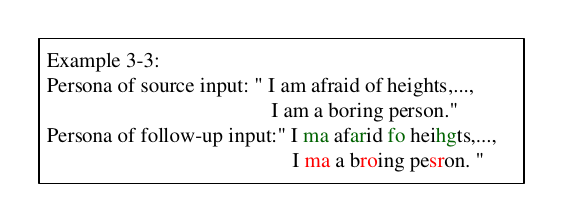}
          \label{MT3-3}
          \vspace{-0.7cm}   
        \end{figure}
        
This paper primarily focuses on metamorphic relations that ensure consistent outputs. However, future work will explore additional relations, including those involving inversed outputs. For example: \emph{Conflicting Persona Information:} It can be referred in \citet{wuda_MT} where Question\&answering tasks are well discussed under metamorphic relations with inversed outputs.

\section{Experiments Setup}

\begin{table*}[]
\centering
\tiny
\resizebox{0.9\linewidth}{!}{
\begin{tabular}{l|c|c|ccccc}
\hline
\multicolumn{1}{c|}{{\color[HTML]{000000} }} &
  {\color[HTML]{000000} } &
  {\color[HTML]{000000} } &
  \multicolumn{5}{c}{{\color[HTML]{000000} MT3}} \\ \cline{4-8} 
\multicolumn{1}{c|}{{\color[HTML]{000000} }} &
  \multirow{-2}{*}{{\color[HTML]{000000} MT1}} &
  \multirow{-2}{*}{{\color[HTML]{000000} MT2}} &
  \multicolumn{1}{c|}{{\color[HTML]{000000} MT3-1}} &
  \multicolumn{2}{c|}{{\color[HTML]{000000} MT3-2}} &
  \multicolumn{2}{c}{{\color[HTML]{000000} MT3-3}} \\ \cline{2-8} 
\multicolumn{1}{c|}{\multirow{-3}{*}{{\color[HTML]{000000} Model}}} &
  {\color[HTML]{000000} Self} &
  {\color[HTML]{000000} Self} &
  \multicolumn{1}{c|}{{\color[HTML]{000000} Self}} &
  {\color[HTML]{000000} Self} &
  \multicolumn{1}{c|}{{\color[HTML]{000000} Partner}} &
  {\color[HTML]{000000} Self} &
  {\color[HTML]{000000} Partner} \\ \hline
{\color[HTML]{000000} DIM (EMNLP 2019)} &
  {\color[HTML]{000000} 25.00\%} &
  {\color[HTML]{000000} 40.43\%} &
  \multicolumn{1}{c|}{{\color[HTML]{000000} 38.19\%}} &
  {\color[HTML]{000000} 41.68\%} &
  \multicolumn{1}{c|}{{\color[HTML]{000000} 65.50\%}} &
  {\color[HTML]{000000} 79.02\%} &
  {\color[HTML]{000000} 68.00\%} \\
{\color[HTML]{000000} FIRE (EMNLP 2020)} &
  {\color[HTML]{000000} 22.26\%} &
  {\color[HTML]{000000} {\ul 38.75\%}} &
  \multicolumn{1}{c|}{{\color[HTML]{000000} 34.66\%}} &
  {\color[HTML]{000000} 41.33\%} &
  \multicolumn{1}{c|}{{\color[HTML]{000000} 61.67\%}} &
  {\color[HTML]{000000} 72.51\%} &
  {\color[HTML]{000000} 64.20\%} \\ \hline
{\color[HTML]{000000} CoBERT (EMNLP 2020)} &
  {\color[HTML]{000000} 24.56\%} &
  {\color[HTML]{000000} 40.87\%} &
  \multicolumn{1}{c|}{{\color[HTML]{000000} 37.33\%}} &
  {\color[HTML]{000000} 43.37\%} &
  \multicolumn{1}{c|}{{\color[HTML]{000000} 67.15\%}} &
  {\color[HTML]{000000} 74.45\%} &
  {\color[HTML]{000000} 71.92\%} \\
{\color[HTML]{000000} BERT\_CRA (SIGIR 2021)} &
  {\color[HTML]{000000} {\ul 20.42\%}} &
  {\color[HTML]{000000} \textbf{38.30\%}} &
  \multicolumn{1}{c|}{{\color[HTML]{000000} {\ul 34.53\%}}} &
  {\color[HTML]{000000} 34.80\%} &
  \multicolumn{1}{c|}{{\color[HTML]{000000} 56.58\%}} &
  {\color[HTML]{000000} 70.71\%} &
  {\color[HTML]{000000} 58.36\%} \\ \hline
{\color[HTML]{000000} prompt\_MLM (IJCNLP 2021)} &
  {\color[HTML]{000000} 20.78\%} &
  {\color[HTML]{000000} 38.90\%} &
  \multicolumn{1}{c|}{{\color[HTML]{000000} \textbf{34.16\%}}} &
  {\color[HTML]{000000} {\ul 33.45\%}} &
  \multicolumn{1}{c|}{{\color[HTML]{000000} {\ul 56.16\%}}} &
  {\color[HTML]{000000} {\ul 69.24\%}} &
  {\color[HTML]{000000} \textbf{57.92\%}} \\
{\color[HTML]{000000} prompt\_NSP (COLING 2022)} &
  {\color[HTML]{000000} \textbf{20.19\%}} &
  {\color[HTML]{000000} 38.94\%} &
  \multicolumn{1}{c|}{{\color[HTML]{000000} 34.94\%}} &
  {\color[HTML]{000000} 35.30\%} &
  \multicolumn{1}{c|}{{\color[HTML]{000000} 56.59\%}} &
  {\color[HTML]{000000} 71.61\%} &
  {\color[HTML]{000000} 60.12\%} \\
{\color[HTML]{000000} prompt-MLM-DialogLM (AAAI 2022)} &
  {\color[HTML]{000000} 33.57\%} &
  {\color[HTML]{000000} 48.42\%} &
  \multicolumn{1}{c|}{{\color[HTML]{000000} 41.01\%}} &
  {\color[HTML]{000000} \textbf{33.15\%}} &
  \multicolumn{1}{c|}{{\color[HTML]{000000} \textbf{54.14\%}}} &
  {\color[HTML]{000000} \textbf{68.57\%}} &
  {\color[HTML]{000000} {\ul 58.27\%}} \\ \hline
\end{tabular}}
\caption{Violation Rate ($V_{r}$) of All metamorphic testing. The models with the lowest violation rates are indicated by bold numbers, while models with the next lowest rates are indicated by underlined numbers. The ``Self" refers to the original self-persona, while the ``Partner" represents the original partner-persona.}
\label{tab:All_test1}
\vspace{-0.4cm}
\end{table*}

\subsection{\textbf{Evaluation Metrics}} \label{metrics}
The assessment of the software outcomes in study \cite{wuda_MT} is quantified as the ``violation rate" (\bm{$V_{r}$}). This metric is employed to gauge the extent to which the models under investigation generate responses that do not satisfy the metamorphic relation. Specifically, we establish the metamorphic relation $MR_{i}\left (t_{i},r\right )$ based on the persona in the personalized dialogue retrieval task, where $t_{i}$ represents a transformation and $r$ denotes the output relation. As previously mentioned, the $r$ signifies consistent output. The test set $\mathbb{S} =\left \{s_{1},s_{2},...,s_{n} \right \}$ is utilized to transform an invariant-example $\mathbb{S^{\mathit{i}}} =\left \{s_{1}^{i},s_{2}^{i},...,s_{n}^{i} \right \}$ based on this relation. Subsequently, we input $\mathbb{S}$ and $\mathbb{S^{\mathit{i}}}$ into the personalized dialogue retrieval model $P$ to obtain results $O_{j}$ and $O_{j}^{i}$, respectively. Based on $r$, we determine that $V_{r}^{i}=1$ if $O_{j}\ne O_{j}^{i} $, and $V_{r}^{i}=0$ otherwise.

For a standardized evaluation of personalized dialogue retrieval models with previous studies,  the recall of true positive responses (\emph{hits@1}) and  mean reverse ranking (\emph{MRR}) are used.


\subsection{Research Questions}
Our study revolves around answering three research questions: 
    
\textbf{RQ1:} \emph{To assess the overall validity of metamorphic testing constructed using persona-based metamorphic relations.}
    As previously discussed, persona plays a vital role in maintaining the consistency of the personality of responses in a multi-turn dialogue model. 
    We have developed persona-centric metamorphic relations to assess the model's robustness based on persona. This experimental question seeks to determine the effectiveness of persona-centric metamorphic testing in identifying flaws in the model under test. It will also evaluate whether our proposed metamorphic relation can achieve the expected advantage.
    
\textbf{RQ2:} \emph{To compare the performance of multi-turn dialogue models trained with different paradigms.}
By conducting metamorphic testing on models that adhere to the three paradigms of non-pretraining, fine-tuning, and prompt learning, and by presenting the test results using the proposed metamorphic relations, we can effectively demonstrate the performance of these models across various facets of their capabilities.
    
\textbf{RQ3:} \emph{To investigate the relationship between metrics based on reference answers and the violation rate metrics of metamorphic testing.}
    In subsection~\ref{metrics}, we will present two metrics utilized in our study: the traditional metric \emph{hits@1} and the violation rate of the metamorphic testing. These metrics evaluate our test results from distinct perspectives. 
    Specifically, we will compare the extent of the decrease in \emph{hits@1} and the percentage of violations reported by the metamorphic testing. We will also assess their error detection effects.

\section{Results and Analysis}

\subsection{RQ1: Persona-based Metamorphic Relations}
Metamorphic testing was conducted on seven multi-turn dialogue models using three persona-centered metamorphic relations, with test cases drawn from \emph{self-persona} and \emph{partner-persona} of the Persona-Chat dataset. Table \ref{tab:All_test1} presents the violation rates observed during testing. The results indicate that all models produced erroneous outputs, as evidenced by violation rates exceeding 0. These findings demonstrate the effectiveness of metamorphic testing in uncovering flaws in the models' abilities to handle persona-related dialogues, and underscore the importance of designing appropriate metamorphic relations for testing such models.

\begin{figure}[ht]
\centering
\includegraphics[width=1\linewidth]{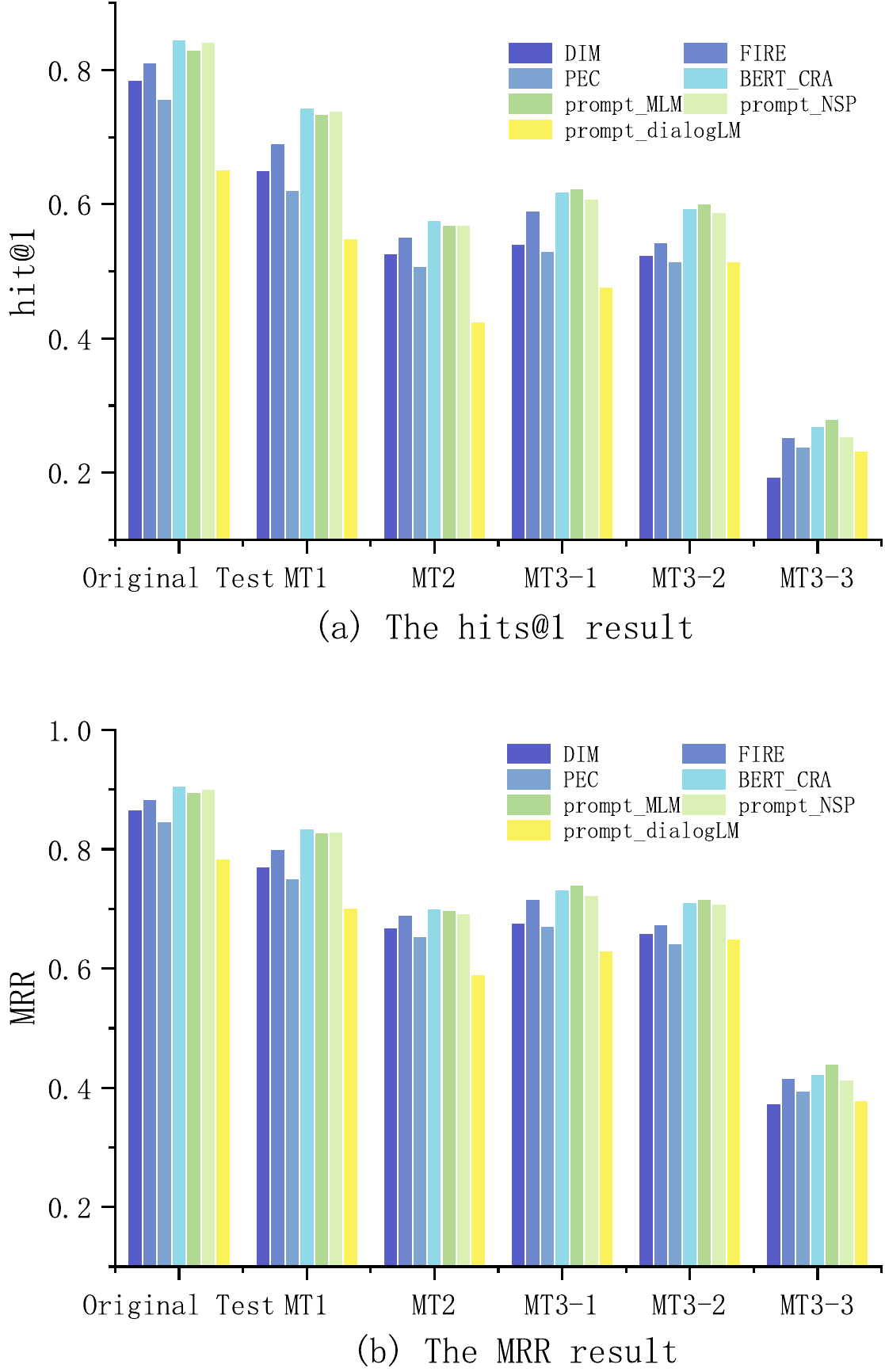}
\caption{The result of \emph{self-persona's original}.}
\label{self-original-recall-mrr}
\vspace{-0.2cm} 
\end{figure}

It is worth noting that we conducted experiments on the last turn of context to compare the impact of noise introduced by persona on personalized dialogue retrieval. Comparing the effects, we observe that the introduction of noise to either persona or context alone has a similar impact on the model, without causing a significant change. However, when noise is simultaneously applied to both persona and context, we observe a sharp increase in the violation rate, nearly twice as much as the effect of applying noise to one aspect alone. Additionally, referencing the significant drop in \emph{hits@1} and \emph{MRR} demonstrated in Figure \ref{self-original-recall-mrr}, this phenomenon suggests that even when the persona and context undergo change separately, the model still possesses the ability to identify a response that aligns with the semantics and personality of the other participant. Hence, it indicates that the model has the capacity to learn and incorporate persona information from the dialogue history. This observation underscores the fact that metamorphic testing can reveal gaps in the models' semantic comprehension abilities.

To summarize, our results indicate that all personalized dialogue retrieval models experienced a decrease in performance when exposed to data perturbation, with the \emph{self-persona} models showing a great vulnerability. 
Therefore, our proposed metamorphic testing approach, along with the application of specific metamorphic relations, shows effective in identifying issues within these models.

\subsection{RQ2: Multi-turn Dialogue Modelling}

The research question we investigate is independent of the dataset used. As Table \ref{tab:All_test1} illustrates, the differences in performance between models in the same scenes are comparable. Therefore, our evaluation primarily centers around the scenes of the \emph{self-persona's original}.

\begin{figure}[ht]
\centering
\includegraphics[width=1\linewidth]{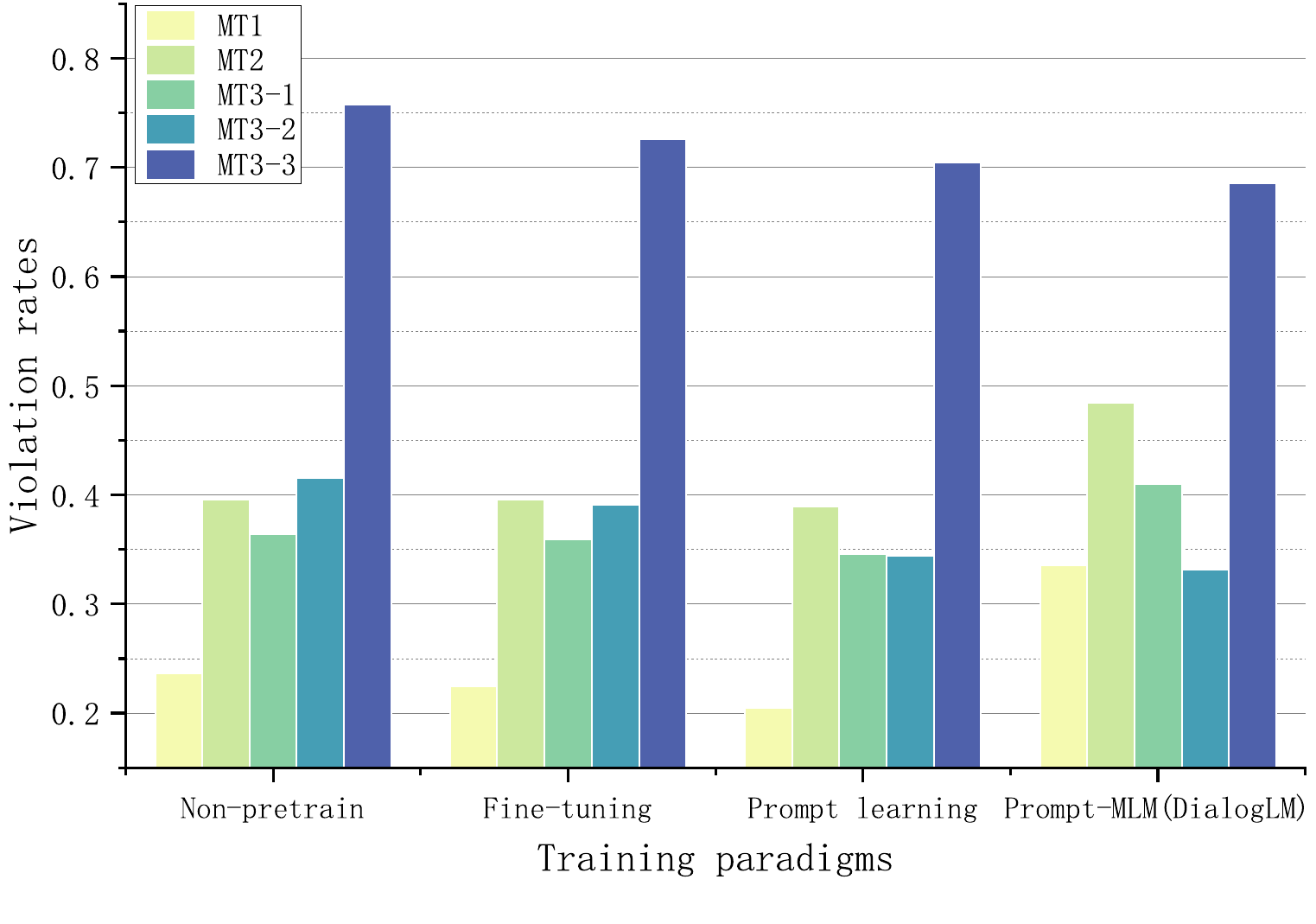}
\caption{The $V_{r}$ result of \emph{self-persona's original} scenario (the average of  the violation rates for their respective included models each paradigm).}
\label{paradigm-vr}
\end{figure}

Figure \ref{paradigm-vr} displays the violation rates of multi-turn dialogue models for the three training paradigms where Seven models are classified into three types. As stated in RQ1, the violation rate for each category exceeds 0, indicating a robustness issue for each model type. Our results reveal that prompt-based models exhibit lower violation rates than those subjected to fine-tune. This aligns with our expectation that prompt learning leverages more latent knowledge acquired during pre-training compared to fine-tune. 

Specially, the prompt-MLM of DialogLM, exhibits the highest violation rate in MT1, surpassing DIM by 8.57\%, which has the second-highest violation rate. This phenomenon can be attributed to DialogLM's pre-training using multi-speaker conversations and long dialog understanding, making our simple prompt setup less suited for its pre-training task compared to BERT. Merely employing prompts without careful consideration not only fails to fully leverage the model's potential knowledge but also hinders its performance, as evidenced by the self-original \emph{hits@1} or \emph{MRR} scores of the prompt-MLM-DialogLM model in Fig \ref{self-original-recall-mrr}.

The violation rates observed in the metamorphic testings based on persona constructions indicate that prompt-based models still possess an advantage in discerning between conversational styles of two speakers based on conversation history, even after introducing perturbations to the personas.

The preceding discussion demonstrates that all model categories experience performance degradation when exposed to data perturbation. 
The robustness of models that incorporate prompts is generally superior to those based on fine-tune, indicating the presence of untapped knowledge in pre-trained models. These findings highlight the need for further research to fully leverage the potential of pre-trained models.

\begin{figure}[ht]
\centering
\includegraphics[width=1\linewidth]{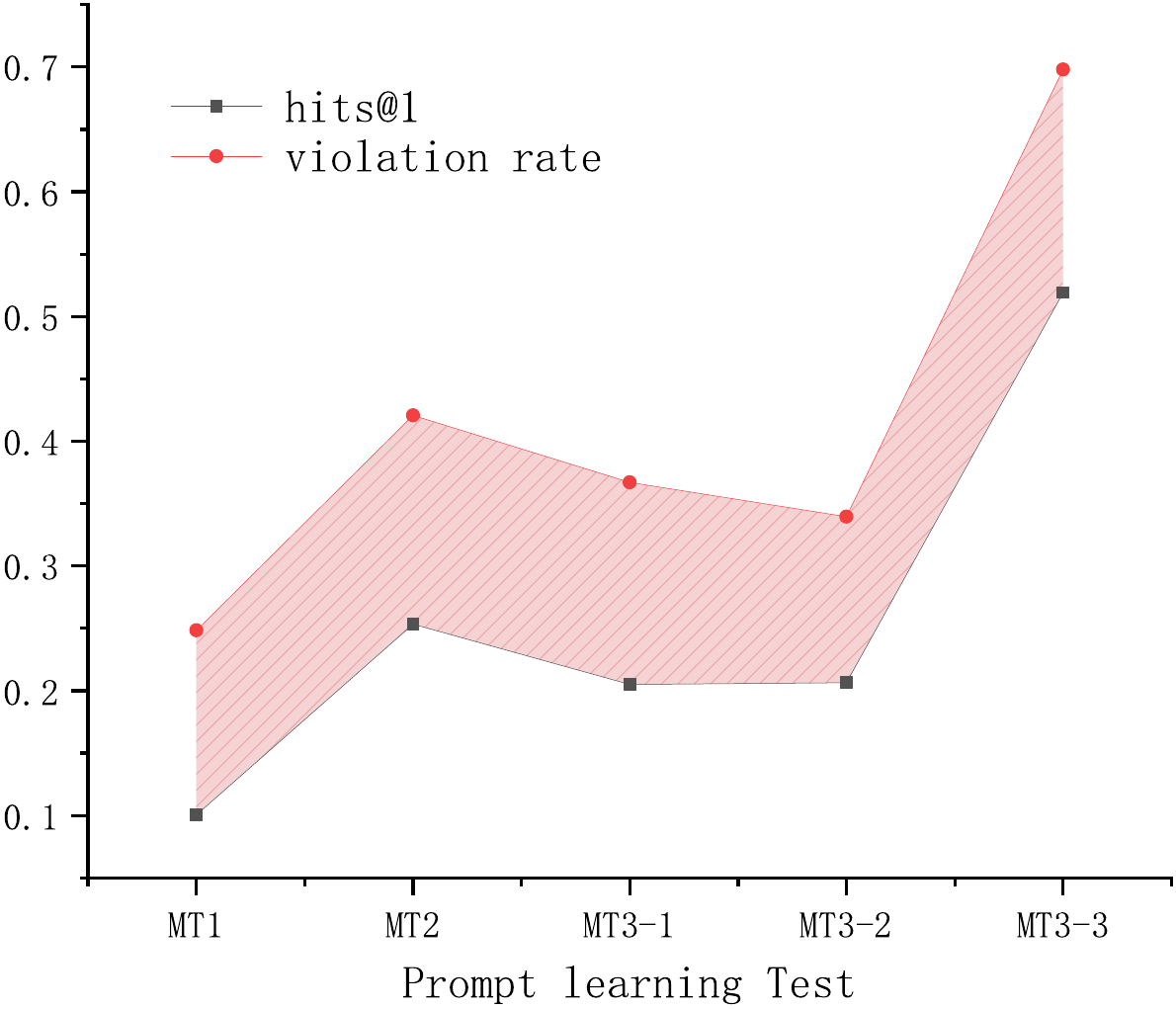}
\caption{Comparison of hits@1 and violation rates($V_{r}$) for prompt learning paradigm. The color region is the part where the violation rate is more than hits@1. Specifically, the red area indicates \emph{self-persona's original}. The values in the figure are the average of the three prompt learning models.}
\label{prompt-hit-vr}
\vspace{-0.2cm} 
\end{figure}

\subsection{RQ3: Reference-Based VS MT}

As Figure \ref{prompt-hit-vr} depicts, the decline in each model's metamorphic testing is relatively limited when evaluated using the \emph{hits@1} metric with labeled answers.  This suggests that these models exhibit a certain level of robustness in terms of synonym recognition, resistance to irrelevant sentence influence, and performance affected by noise. However, it is important to note that this approach, which employs standardized answers for testing, can be considered a type of adversarial testing. Nonetheless, it doesn't provide sufficient insight into the internal deficiencies of the models or their actual understanding of the required knowledge, like the limitations described in Sec.~\ref{limitation}.

On the other hand, by utilizing $V_{r}$ to assess the results of the model test, we were able to identify more errors in the models. Moreover, the use of $V_{r}$ highlights the possibility of answerless scenarios in metamorphic testing. By establishing a metamorphic relation based on persona consistency and constructing metamorphic testing, we gain a comparative advantage over adversarial testing. 
During the analysis of the preceding two experimental questions (RQs), we leveraged the traditional \emph{hits@1} metric to support the evaluation of violation rates in personalized dialogue retrieval tasks with labeled answers. This approach enabled us to conduct a more thorough examination of anomalies in the MT results.

\subsection{More Discussions}

Our study introduces \textbf{three specific metamorphic testings} to examine persona-based output consistency in personalized dialogue retrieval. 
The results effectively demonstrate the utility of our work in identifying potential comprehension issues within models. 
We observe significant performance degradation across different model types when exposed to intentional noise, emphasizing the critical role of our tests in revealing limitations.
The intentionally introduced noise represents an extreme scenario, providing a lower-bound estimate of the model's performance in real-world scenarios with natural noise. 
Our analysis reveals the model's proficiency in grasping context and persona characteristics, successfully matching response to specific persona even with partial information loss.

Our evaluation of \textbf{three training paradigms} reveals that fine-tuning and prompt learning through pre-training exhibit greater stability than non-pretrained models, showing less susceptibility to data perturbations and performance degradation. 
This enhanced stability can be attributed to the fact that pre-trained models are trained on large-scale corpora, allowing them to better handle noise and improve generalization. 
Prompt learning effectively leverages the latent knowledge of the pre-trained model, but constructing templates within the pre-trained format does not fully unleash its capability and can lead to performance degradation. 
In summary, existing personalized dialogue retrieval models have room for improvement in performance.

Hybrid retrieval and generative modules are commonly used in dialogue systems, leading us to investigate \textbf{the robustness of generative models}.  
Additionally, we find that the persona-centric metamorphic relation in personalization-based dialogue systems is also applicable to generative models. 
For example, in Sec.~\ref{MR1}, we expect a robust generative model to maintain the originality of generated sentences when introducing persona sentences with similar meanings.

\section{Related Work}

\subsection{Reference-based Validation on Multi-turn Dialogue System}
The concept of persona was introduced by the researchers to ensure consistent characterization and clear memory for generating rational responses \cite{persona-chat_dataset,PEC}.
Evaluation of personalized dialogue systems commonly involves a reference-based approach to ensure persona consistency. Standardized labels, generated through human evaluation, are used alongside traditional metrics like \emph{hits@1} to assess system performance \cite{persona-chat_dataset,BERT-CRA,COLING,TransferTransfo}. This ensures robust and accurate measurement of system effectiveness.
Some researchers \cite{43} have explored robustness in retrieval-based dialogue systems by generating adversarial examples in black-box settings, still the continued need for labeled data.

Metamorphic testing 
complements traditional reference-based metrics, focusing on targeted assessment. An important benefit is its reliance on existing annotated labels, enabling automated generation of additional testing data without further manual annotation.

\subsection{Metamorphic Testing In NLP}
Metamorphic testing \cite{11} is an approach used to evaluate the internal consistency of NLP models by examining the preservation of expected relationships between inputs and outputs \cite{Beyond_Accuracy}. 
A significant focus of existing metamorphic relations in NLP is to assess the robustness of NLP models. These relations specifically evaluate the ability of a model to maintain consistent output \cite{Stress_Test,15,17,Assessing}.
Demonstrating their effectiveness, robustness relations have been successfully applied in testing various NLP tasks, including Sentiment Analysis \cite{Beyond_Accuracy,18} and NLI \cite{Stress_Test,Assessing}, among others.
Additionally, metamorphic testing has been extended to test other aspects of model performance, including fairness \cite{Fairness_Violations}, and more\cite{23,Systematicity_Perspective}. 
This highlights the versatility of metamorphic testing in assessing different dimensions of NLP models.

Metamorphic testing in NLP is being explored in various tasks with mixed success. 
Our work aims further to investigate persona-based consistent outputs to assess its robustness with the help of MT.

\section{Conclusions}

In this paper, we conduct an evaluation of three paradigms using three persona-centric relational construct tests, aimed at uncovering the strengths and weaknesses of personalized dialogue models in a comprehensive manner. Our findings indicate that all of the models experience performance degradation, highlighting that their accuracy performance alone does not necessarily reflect their actual effectiveness, as they are susceptible to data perturbations. Notably, the prompt learning paradigm exhibits relatively higher stability, but we also discover that the design of the template significantly influences the model's stability. 

\section{Limitations}
Although our proposed persona-centric metamorphic relation partially uncovers potential issues with personalized conversation retrieval models, there are remaining considerations for generation based dialogue system. Our designed metamorphic relations can be applied to other dialogue modelling besides retrieval based one. Further extension and validation of our metamorphic relations on relevant conversation datasets are needed. Moving forward, we aim to expand our evaluation by incorporating additional metamorphic relations such as considering fairness and biases. Furthermore, we plan to conduct assessments on some specific dialogue datasets, including those designed to explore empathic dialogues. 

\nocite{*}
\section{Bibliographical References}\label{sec:reference}

\bibliographystyle{plainnat}
\bibliography{lrec-coling2024-example}


\end{document}